\documentclass{aastex}   
\usepackage{emulateapj5}
    
\slugcomment{submitted to \apj}

\shorttitle{Heating and conduction in cooling flows}
\shortauthors{M. Ruszkowski and M.C. Begelman}

\begin{document}

\title{Heating, conduction and minimum temperatures in cooling flows}

\author{Mateusz Ruszkowski and Mitchell C. Begelman\altaffilmark{1}}
\affil{JILA, Campus Box 440, University of Colorado, Boulder CO 80309-0440}
\email{mr@quixote.colorado.edu; mitch@jila.colorado.edu}

\altaffiltext{1}{Also at Department of Astrophysical and Planetary
Sciences, University of Colorado}

\begin{abstract}
There is mounting observational evidence from {\it Chandra} for strong
interaction between keV gas and AGN in cooling flows. It is now widely
accepted that the temperatures of cluster cores are maintained at a level
of $\sim$ 1 keV and that the mass deposition rates are lower than earlier
{\it ROSAT/Einstein} values. Recent theoretical results suggest that
thermal conduction can be very efficient even in magnetized plasmas.
Motivated by these discoveries, we consider 
a ``double heating model'' which incorporates
the effects of {\it
simultaneous} heating by both
the central AGN and thermal conduction from the hot outer layers of 
clusters. Using hydrodynamical simulations, we demonstrate that there 
exists a family of solutions that does not suffer from the cooling
catastrophe. In these cases, clusters relax to a stable final state,
which is characterized by minimum temperatures of order 1 keV and
density and temperature profiles consistent with observations.
Moreover, the accretion rates are much reduced, thereby reducing the need
for excessive mass deposition rates required by the standard cooling flow
models. 

\end{abstract}


\section{Introduction}
Radiative cooling of gas in the central regions of galaxy clusters 
occurs on a timescale much
shorter than the Hubble time. The cooling time-scale of this gas
increases with distance from the cluster core. In the absence of any
heating sources, this implies that the intracluster medium must
accrete subsonically 
toward the center in order to maintain pressure equilibrium with the 
gas at larger radii. The mass deposition rates predicted by
this cooling flow model are very high and range typically 
from 10 to 1000 solar masses
per year. X-ray observations made prior to the launch of the 
{\it Chandra Observatory} seemed to be 
broadly consistent with this picture \citep{fa94}.
Although both the gas temperature and cooling time are observed to decline
toward cluster cores, new {\it Chandra} and {\it XMM-Newton}
 observations show a
remarkable lack of emission lines from gas at temperatures below $\sim 1$
keV in the central regions of clusters \citep{pe01,al01}.
Moreover, the cooling mass deposition rates obtained with {\it Chandra} and
{\it XMM} 
using spectroscopic methods are $\sim 10$ times smaller than earlier
estimates based on {\it ROSAT} and {\it Einstein} observations
\citep{mc01,da01,pe01}. 
On the other hand, morphological cooling rates 
give accretion rates vastly exceeding the ones based on
spectroscopic methods \citep{da01}. 
The strong discrepancy between these results 
indicates either that the gas is prevented from cooling
by some heating process or that it cools without any spectroscopic
signatures \citep{fa01}.\\ 
\indent
In this paper, we consider cooling
flow models that incorporate the effects of heating by central active
galactic nuclei and thermal conduction from the hot outer layers of
clusters. We show that evolving density and temperature profiles can relax
to stable final states, which are consistent with X-ray observations.
Equilibria are characterized by minimum temperatures of order 1 keV at
radii $\sim$ 1 kpc.
The paper is organized as follows. In section 2 we discuss
the observational and theoretical motivation for including both AGN heating and
thermal conduction. In section 3 we present the details of our model. 
The results of hydrodynamic simulations are presented in Section 4 and the
main results are summarized in section 5.

\section{Heating mechanisms}
\subsection{Feedback and heating by AGN}
A large fraction ($\sim 70\%$) of cD galaxies in the centers of cooling
flow clusters shows evidence for powerful radio activity. 
This fraction is much larger than in non-cooling flow clusters
and suggests a link between the presence of the cooling gas and the
activity of the central supermassive black hole.
Currently, {\it Chandra} offers the best opportunity to study this 
coupling as the cluster cores are optically thin and can be
spatially resolved
down to scales of a few kpc within $z<0.1$. These scales are 
comparable to the sizes of the radio sources.
Recent {\it Chandra} observations of the Perseus, Hydra A and many other
clusters reveal holes in the X-ray surface
brightness which coincide with radio lobes \citep{fa01,mc01}.
These radio sources inflate bubbles of hot
plasma that subsequently rise through the cluster atmosphere, mixing and 
heating it up. This mechanism has been invoked to explain the reduced
cooling rates \citep{jo01,ch02}.

\subsection{Heating by thermal conduction}
A potential difficulty with reducing the cooling accretion rate by
means of a central heating source is that too strong a heating source will lead
to very strong convection, which could remove metallicity gradients
observed by {\it Chandra} in cluster cores \citep{joh02}.
Therefore, the limits on energy requirements of central AGN  
and on the amount of mixing and convection suggest that an additional
heating source is required to offset cooling rates.
An important candidate is thermal conduction
of heat from the outer hot layers of the cooling flow cluster. 
With some exceptions \citep{bm86,ma01,bm02,fv02,fvm02}
this effect has been largely neglected in
observational and theoretical studies due to the
preconception that the conduction coefficient would be suppressed much
below the classical Spitzer value by magnetic fields. 
Indeed, there is 
strong evidence that clusters are magnetized by intermittent radio galaxies
in cluster cores \citep{cl01,kr01,rb97}.
However, recent theoretical work by \citep{nm01} suggests that
if the magnetic field is highly turbulent, then thermal conduction is
relatively 
efficient. They find that if the turbulence spectrum extends over two or more
decades in wave vector, then the 
thermal conduction coefficient is only a factor $\sim $ a few
below the Spitzer value and thus could play a significant role in cooling
flows in clusters of galaxies.
The other argument in favor of thermal conduction is that
it is a very strongly increasing
function of temperature and, therefore, could be efficient in the
outer layers of clusters where temperatures are high.\\

\subsection{Simultaneous heating by AGN and conduction}
If radiative cooling is balanced solely by energy input 
from a central supermassive black hole then a cooling flow can be 
quenched provided that a sufficient amount of power is provided in
kinetic form \citep{mc02}. 
Models based on the assumption that radiative cooling is balanced by energy
input from AGN
predict that the mechanical power of AGN in cooling flows is much higher than the presently
observed bolometric luminosities of these objects \citep{ch02}.
However, the 
power necessary to offset a cooling flow may still be present in the
kinetic form after the central AGN becomes inactive. In fact, the required
power is consistent with estimates of jet power in some objects.
Theoretical models of impulsive central heating in elliptical galaxies 
\citep{bt95} predict violent successive cooling catastrophes and temperature rising towards
the center for radii less than $\sim 100$ kpc.
On the other hand, attempts to build cooling flow models with conduction as the only heating
mechanism fail. 
Such models predict that the cooling time of the central cluster region
diminishes to less than the free-fall time, leading to a cooling
catastrophe \citep{me88}. This results in supersonic accretion
of cold and dense material in the core, which is contrary to observations. 
Stable models can be obtained but they assume a significant amount of
distributed mass dropout
which does not have observational grounding. 
Moreover, even when the models are stable,
such cooling flows suffer from additional problems.
Due to the strong dependence of the conduction coefficient on
temperature ($\propto T^{5/2}$), the predicted temperatures either 
drop by less than a factor of 2 or the central temperatures are so low that 
heat conduction is negligible and the central accretion rate 
often becomes very high. This suggests that
heat conduction alone cannot substantially reduce mass accretion rates
everywhere within a cluster. 
Recently, \citep{bm02} considered centrally 
heated cooling flow models with conduction.
They concluded that such models are grossly
incompatible with observations and were unable to find 
even a single acceptable
cooling flow model using their heating prescriptions.\\
\indent
In the next section we present details of our model and look at 
the consequences of bridging these two heating regimes. 

\section{``Double heating'' model}
\subsection{Physical assumptions}
We solve numerically the equations of hydrodynamics in the following form

\begin{eqnarray}
\frac{\partial\rho}{\partial t}+\nabla\cdot(\rho\mathbf{v}) & = & 0\\
\frac{\partial\mathbf{S}}{\partial
t}+\nabla\cdot(\mathbf{Sv}) & = & -\nabla p -\rho\nabla\Psi\\
\frac{\partial e}{\partial
t}+\nabla\cdot(e\mathbf{v}) & = & -p\nabla\cdot\mathbf{v} 
-\nabla\cdot\mathbf{F}_{\rm cond}-\nabla\cdot\mathbf{F}_{\rm conv} \nonumber \\
 & & -n_{e}^{2}\Lambda (T)+\cal{H},
\end{eqnarray}

\noindent
where $\rho$ is the mass density, $n_{e}$ the electron number density, 
$p$ the pressure, $\mathbf{v}$ the velocity,
$\mathbf{S}=\rho\mathbf{v}$ the momentum vector,
$e$ the internal energy density, $\Psi$ the gravitational potential,
$\Lambda(T)$ the cooling function and $\cal{H}$ the heating rate per
unit volume. We adopt an equation of state $p=(\gamma -1)e$ and
consider models with $\gamma =5/3$. The conductive flux $\mathbf{F}_{\rm cond}$ is
given by

\begin{equation}
\mathbf{F}_{\rm cond}=-f\kappa T^{5/2}\nabla T,
\end{equation}

\noindent
where $\kappa$ is the Spitzer conductivity 

\begin{equation}
\kappa =\frac{1.84\times 10^{-5}T^{5/2}}{\ln\lambda},
\end{equation}

\noindent
with the Coulomb logarithm $\ln\lambda = 37$, and 
$f$ is a reduction factor $(0\le f\le 1)$. 
The saturation of the conductive flux was not important for the parameters
considered in this paper.
The convective flux $\mathbf{F}_{\rm conv}$ is given by mixing length theory

\begin{equation}
\mathbf{F}_{\rm conv} =\left\{ \begin{array}{ll}
\frac{1}{2^{5/2}c_{p}}g^{1/2}\rho l^{2}_{m}(-\nabla \hat{s})^{3/2} & \mbox{if
$\nabla \hat{s}<0$}\\
0 & \mbox{otherwise}
\end{array}\right. ,
\end{equation}

\noindent
where $g$ is the gravitational acceleration, $l_{m}$ is the
mixing length, $\hat{s}=c_{v}\ln [(\gamma -1)e/\rho^{\gamma}]$
is the gas entropy, $c_{v}$ is the specific heat per unit volume, 
$c_{p}=\gamma k_{B}/[(\gamma-1)\mu m_{p}]$ is the
specific heat per unit mass at constant pressure, 
$\mu =0.5$ is the mean molecular weight and $\gamma$ is the adiabatic index.
In mixing length theory $l_{m}$ is a free parameter. 
We use $l_{m}=\min [0.3(\gamma -1)e/(\rho g),r]$, where $r$ is the distance from the cluster center.
Motivated by recent {\it Chandra} X-ray observations of clusters
(e.g., \citep{sc01,al01}), we parameterize the dark matter distribution using 
a non-singular isothermal mass density profile. We add a
contribution from the central galaxy also in the form of a non-singular
isothermal profile but characterized by different values of parameters.
We assume that the total potential does
not evolve. The contribution of the gas to gravity is negligible throughout
the simulation volume.
The corresponding total gravitational potential 
is then given by $\Psi_{\rm tot} =\Psi_{c}+\Psi_{g}$, where

\begin{equation}
\Psi_{c,g}=\sigma_{c,g}^{2}\ln \left[1+\left(\frac{r}{r_{c,g}}\right)^{2}\right]+
2\sigma_{c,g}^{2}\left(\frac{r_{c,g}}{r}\right)\arctan
\left(\frac{r}{r_{c,g}}\right),
\end{equation}
where $\sigma_{c,g}$ is the cluster or galactic velocity dispersion and
 $r_{c,g}$ is the core radius of the cluster or galaxy.\\
\indent
Note that equations (1)-(3) do not include terms related to mass dropout rate.
Such terms are usually invoked to avoid a cooling catastrophe by removing
cooled gas from the flow. 
Thermal instabilities, which lead to mass dropout, appear when the 
cooling time is comparable to or shorter than the dynamical time. 
In our simulations the cooling time is longer by a factor $10 -
 100$ than the dynamical time, so our assumption is justified.
A small amount of mass dropout may still occur in the central regions if
the gas becomes locally overdense, e.g., due to interactions of the 
central AGN with the surrounding medium.  
However, as mentioned above, there is no strong
observational indication that significant amounts of gas are decoupling throughout
the cooling flow region, although star-forming regions and 
small amounts of cool gas in the form of
 $H\alpha$ emission have been detected in the innermost
 regions of clusters. Such colder gas also could have been lifted from 
very small radii to larger distances by the central AGN.

\subsubsection{Heating and cooling}
Following \citep{to01} we use an approximation to the cooling function
based on detailed calculations by \citep{sd93} 

\begin{equation}
n_{e}^{2}\Lambda =[C_{1}(k_{B}T)^{\alpha}+C_{2}(k_{B}T)^{\beta}+C_{3}]n_{i}n_{e},
\end{equation}

\noindent
where $n_{i}$ is the ion number density and the units for $k_{B}T$ are $keV$. 
For an average metallicity $Z=0.3
Z_{\odot}$ the constants in equation (8) are $\alpha =-1.7$, $\beta =0.5$, $C_{1}=8.6\times
10^{-3}$, $C_{2}=5.8\times 10^{-2}$ and $C_{3}=6.4\times 10^{-2}$ and we
can approximate $n_{i}n_{e}=0.704(\rho/m_{p})^{2}$. 
The units of $\Lambda$ are $10^{-22}$ erg cm$^{3}$ $s^{-1}$.
This cooling function
incorporates the effects of free-free and line cooling.\\
\indent
The final term in equation (3) represents heating by the central AGN.
Since it is ultimately heating by mechanical energy and particles from
radio sources, it will be distributed in radius.
We adopt a physically motivated ``effervescent heating'' mechanism \citep{be01} to describe 
the volume heating rate. 
The physical motivation for this model is as follows.
Suppose the central radio source deposits buoyant gas which distributes
itself relatively evenly among bubbles or filaments but does not mix
microscopically with the intracluster medium (ICM). These bubbles will then rise
through the ICM and expand because of the non-negligible pressure
gradient. Bubble expansion will be associated with the conversion of the internal
bubble energy to kinetic form and, eventually, to
heat due to disorganized motion of the ICM.
Since the bubbles rise on a timescale shorter than the radiative cooling
time, the mechanism should reach a quasi-steady state and the dependence on
details of bubble filling factor, rise rate, etc. cancel out.
In a steady state (and assuming spherical symmetry), the energy flux
available for heating is then

\begin{equation}
\dot{e}\propto p_{b}(r)^{(\gamma_{b}-1)/\gamma_{b}},
\end{equation}

\noindent
where $p_{b}(r)$ is the partial pressure of buoyant gas inside bubbles at
radius $r$ and $\gamma_{b}$ is the adiabatic index of buoyant gas. 
Assuming that the partial pressure scales with the thermal
pressure of the ICM, the volume heating function $\cal{H}$ can be expressed
as 

\begin{equation}
\mathcal{H}\sim -h(r)\nabla\cdot\frac{\dot{e}}{4\pi
r^{2}}=-h(r)\left(\frac{p}{p_{0}}\right)^{(\gamma_{b} -1)/\gamma_{b}}\frac{1}{r}\frac{d\ln p}{d\ln r},
\end{equation}

\noindent
where $p_{0}$ is the central pressure and $h(r)$ is the normalization function

\begin{equation}
h(r)=\frac{L}{4\pi r^{2}}(1-e^{-r/r_{0}})q^{-1}.
\end{equation}

\noindent
In equation (11), $L$ is the luminosity of the central source and $q$ is defined by

\begin{equation}
q=\int_{r_{\rm min}}^{r_{\rm max}}\left(\frac{p}{p_{0}}\right)^{(\gamma_{b} -1)/\gamma_{b}}\frac{1}{r}\frac{d\ln p}{d\ln r}(1-e^{-r/r_{0}})dr,
\end{equation}

\noindent
where $r_{0}$ is the inner heating cutoff radius.
Interestingly, the cosmic ray heating model for cooling flows \citep{lo91} predicts
a very similar functional form for the heating function. In this model,
heating is due to cosmic rays, which are produced in the central active nucleus
and are trapped by Alfv\'en waves in the cooling flow plasma. 
They also do work on the thermal gas as they propagate down the pressure gradient.
Radio galaxies are likely to be intermittent on a  
time scale much shorter than the Hubble time, and possibly as short as
$t_{i}\sim 10^{4}-10^{5}$ yr \citep{rb97}. 
The bubble rise speed would be comparable to the sound speed,
which in turn is comparable to the dynamical speed. Therefore, the cooling
timescale is much longer than the bubble rise timescale, and 
it is justifiable to treat feedback heating $L$ in a time-averaged sense.
Therefore, in equation (11) we assume that the luminosity is injected
instantaneously and neglect any delay between central activity and heating
of the ICM.
We also assume that all energy generated in the center goes into
heating. In principle some fraction of the energy could escape the cluster in the
form of sound waves. 
However, they are likely to carry away only a fraction of the energy
comparable to that which is dissipated (e.g., \citep{rey02,ch02}).
The physical motivation for the inner heating cutoff arises from 
the finite size of the central radio source. 
Energy from such a source will start to dissipate at a finite distance from the
cluster center.
Unlike heating functions that depend on local microscopic physics, this
heating function is nonlocal in the sense that it depends on the pressure
gradient. In this regard, it resembles thermal conduction, but here
the heating rate depends on the gradient of pressure rather than
temperature. For simplicity and in order to avoid numerical complications, 
we adopt hydrostatic values to calculate the logarithmic derivative in the
code (i.e., $\nabla p=-\rho\nabla\Psi$).\\
\indent
In the specific case considered below, the feedback luminosity is assumed to be
$L\sim-\epsilon\dot{M}c^{2}$, where $\dot{M}=4\pi r^{2}_{\rm min}\rho v$ is the accretion rate
at the inner radius $r_{\rm min}$ and $\epsilon$ is the accretion efficiency.

\subsection{Numerical methods}
All of the calculations presented in this paper use the ZEUS-3D
\citep{cnf94} code in its 1D mode. The code has been modified by including
a non-singular isothermal potential, feedback heating, cooling, convection 
and conduction. For stability, the convection and conduction terms have to
be integrated using time steps that satisfy appropriate Courant
conditions. The Courant conditions for conduction and convection read

\begin{equation}
\Delta t_{\rm conv}\le\frac{1}{2}\frac{e(\Delta r)^{2}}{f\kappa T^{5/2}}
\end{equation}
\begin{equation}
\Delta t_{\rm cond}\le\frac{1}{2}\frac{2^{5/2}}{\gamma}\frac{(\Delta
r)^{5/2}}{g^{1/2}l_{m}^{2}} .
\end{equation}

\noindent
Whenever either (or both) of these timescales is smaller than the time step
used for hydrodynamic equations, 
we integrate the respective terms separately at the smaller time step.
This method is analogous to the one used by \citep{spb99} in hydrodynamical
simulations of viscous non-radiative accretion discs.\\
\indent
Our computational grid extends from $r_{\rm min}=1$ kpc to $r_{\rm
max}=200$ kpc. In order to resolve adequately the inner regions it is
necessary to adopt a non-uniform grid. We use a logarithmic grid in which 
$(\Delta r)_{i+1}/(\Delta r)_{i}=\sqrt[N-1]{10}$, where $N$ is the number of
grid points. Our standard resolution is $N=400$.

\subsubsection{Initial and boundary conditions}
In order to generate initial conditions we solve the equation of hydrostatic
equilibrium of a gas at constant temperature. 
We assume that the gas
density initially constitutes a fraction $f_{\rm igas}$ of the cluster 
dark matter density at the inner
radius. The gas is assumed to be in
contact with a pressure and thermal bath at the outer radius. Thus, we ensure that
temperature and density at the outer radius are constant.
We adopt inflow/outflow
boundary conditions at the outer radius but at the inner boundary we use
outflow boundary conditions. 
We extrapolate hydrodynamic variables from the active
zones to the ``ghost'' zones used to compute the derivatives of the hydrodynamic variables.
At the inner boundary we allow mass to flow out of 
the computational region ($v\le 0$) and use a switch to ensure that the 
fluid can only flow toward the center (i.e., 
boundary value of the velocity is
zero if the gas tries to flow in the opposite direction). Feedback is
assumed to be present only when the gas at the inner boundary flows inwards
($v<0$).

\section{Results}
To illustrate our method we now present one representative model.
This model reproduces the main features of the observed cooling flows including
a floor in the temperature at about 1 keV. We note that 
fine tuning is not required to obtain stable equilibrium solutions.
Figure 1 shows the evolution of temperature (upper left panel), electron
number density (upper
right panel), entropy ($S\equiv k_{B}T/n_{e}^{\gamma -1}$; 
lower left panel) and mass
accretion rate as a function of distance from the cluster center.
We follow cluster evolution for one Hubble time $t_{H}=H_{0}^{-1}\;
(H_{0}=75\; {\rm km}{\rm s}^{-1}{\rm Mpc}^{-1})$. Each curve within a given
panel is plotted every 
$1/40$ or $1/100$ of the Hubble time, as specified in the caption.
The initial gas temperature is set to be 4 keV throughout the
cluster. 
Specific parameters of the model presented in Figure 1 are as follows:
$\epsilon=0.003$, $\gamma_{b}=4/3$, $\gamma=5/3$, $r_{0}=20
\; {\rm kpc}$, $r_{c}=30\; {\rm kpc}$, $r_{g}=4\; {\rm kpc}$,
$\sigma_{g} =800\;{\rm km}\; s^{-1}$, 
$\sigma_{c} =300\;{\rm km}\; s^{-1}$,
$f=0.23$, $Z=0.3 Z_{\odot}$ and $f_{igas}=0.025$ 
(which corresponds to the final gas mass fraction
$f_{\rm gas}(r<200\; {\rm kpc})=M_{\rm gas}(<r)/M_{\rm tot}(<r)\sim 0.06$). \\
\indent
Line and free-free cooling in the cluster center leads to a 
slow decrease in temperature. At this stage feedback is not yet very
important. The initial phase of slow cooling is followed by a gradual increase in 
cooling rate, which is caused mainly by the increase of density in the
center. The feedback heating is controlled by the accretion rate in the center.
Therefore, the cluster does not cool in runaway fashion.
Once the average inflow velocity in the center has become sufficiently high
and sufficient density has accumulated in the center of the cluster,
the feedback becomes strong enough to suppress further decrease in temperature.
The cluster then quickly relaxes to an equilibrium state. 
The evolution of gas density is similar. Initially, gas  
accumulates gradually in the center of the cluster. When feedback becomes
strong, further accumulation is prevented and the central density
is stabilized. Entropy and mass accretion rate exhibit a similar behavior -- 
the former relaxes to a stable equilibrium and the latter, after
oscillations through positive and negative values, tends to a
small constant negative value as a function of radius.
In the final state, the accretion rate at the inner radius is $\dot{M}\sim
-1.76M_{\odot}$ yr$^{-1}$, 
which is much smaller than acretion rates inferred from
the standard cooling flow models.
Our $\dot{M}$ corresponds to the bolometric feedback luminosity
$L\sim 3\times 10^{44}$ erg s$^{-1}$. 
The most important aspect of the cluster evolution is that it is possible 
for the cluster to reach sustainable and stable equilibrium. In our model, 
there is no need for mass dropout distributed throughout the cooling flow 
region. However, some fraction of the material in the central regions 
may become thermally unstable and form H$\alpha$ emitting filaments or
stars (e.g., \citep{bla01,mc02}). 
Adding mass dropout terms corresponding to these effects would
only improve the stability properties of our models. It could also decrease
the central gas density and, thus, lead to a higher minimum temperature.
\\
\indent
Figure 2 shows the temperature, electron number density, entropy and mass
accretion rate as a function of time (in units of Hubble time
$t_{H}=H_{0}^{-1}$, $H_{0}=75$ km s$^{-1}$ Mpc$^{-1}$) for
different distances from the cluster center $r=5,\; 10,\; 20,\; 40,\; 80,\; 160$ kpc. 
The parameters are the same as for Figure 1, i.e., 
Figure 2 shows cross-sections through the panels in Figure 1 at the
above radii. In the case of entropy
and temperature, the above sequence of $r$
corresponds to the curves from bottom to top.
For density the trend is opposite. In the case of accretion rate, the
amplitude of oscillations increases with 
$r$ (= $5,\; 10,\; 20,\; 40,\; 80$). 
The oscillations are due to sound waves, which propagate
across the cluster as it adjusts to changing conditions.
The precise character of these sound waves, i.e., number of peaks 
per unit time, depends on the boundary conditions.
As can be clearly seen, the
slow initial evolution of the cluster is followed by a faster ``collapse''
stage around $t=0.2t_{H}$. After this phase, the cluster stabilizes
and reaches its final equilibrium state. The profiles shown in Figure 1 
bear close resemblance to the observed temperature, density and
entropy profiles (e.g., \citep{joh02} {\it Chandra} observations of Abell
2199).\\
\indent
Although nominally present in the code, convection is not important for the
parameters of the model presented in this paper. The intracluster medium
usually does not become convectively unstable. We also considered
models with stronger heating in the center (i.e., higher efficiency
$\epsilon$). In such models the gradient of
entropy was significantly negative and convection was present. However, strong heating
prevented the gas from accreting in the center. This led to the
accumulation of material and, subsequently, to a cooling catastrophe.\\
\indent

\section{Conclusions}
We have proposed a new class of time-dependent cooling flow models where cooling is
offset by a combination of central AGN heating and thermal conduction from the
outer regions.
 Our models do not require any mass dropout rate distributed throughout the
cluster. 
We showed that it is possible to obtain stable final equilibrium states, which
do not suffer from the cooling catastrophe.
We have presented a representative model, which reproduces
the main features of observed cooling flows, including a floor in
the temperature at about 1 keV. We have also found stable models for other
parameters, which we will present later.
Fine tuning is apparently not required to
obtain stable equilibrium solutions.
Moreover, stable models are characterized by gas accretion rates that are
 much smaller than
the mass dropout rates predicted by standard cooling flow models.\\

We are grateful to Phil Armitage, Fabian Heitsch, Christian Kaiser
and Daniel Proga for helpful discussions and the referee for a fast response. 
This work was supported in part by NSF grant AST--9876887.

\newpage
\begin{figure*}
\begin{center}
\includegraphics[angle=0,width=0.8\textwidth,height=0.45\textheight]{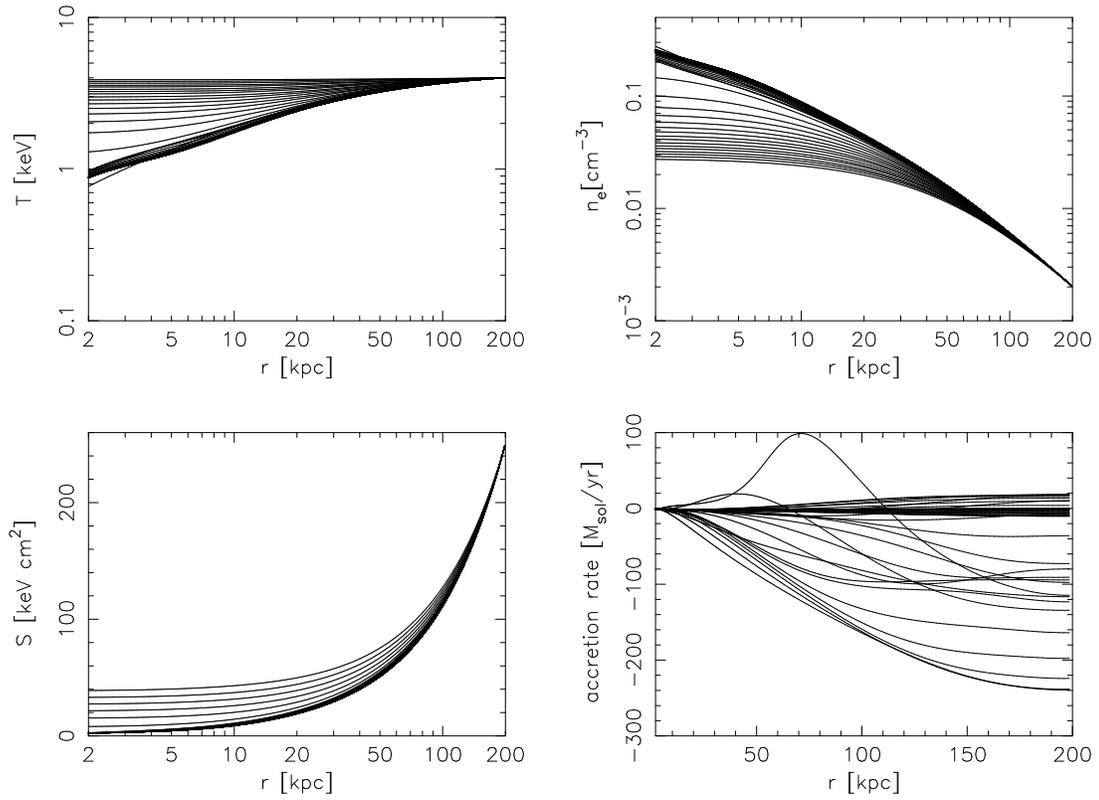}
\caption{Time sequence of temperature (upper left panel), electron number density (upper
right), entropy ($S\equiv k_{B}T/n_{e}^{\gamma -1}$, lower left) and
accretion rate. Temperature and density are shown every $0.01H_{0}^{-1}$ 
and entropy and accretion rate every
$0.025H_{0}^{-1}$. The model settles down to a stable equilibrium state,
which is visible via the dense concentration of curves. See text for additional information.}
\end{center}
\end{figure*}

\begin{figure*}
\begin{center}
\includegraphics[angle=0,width=0.8\textwidth,height=0.45\textheight]{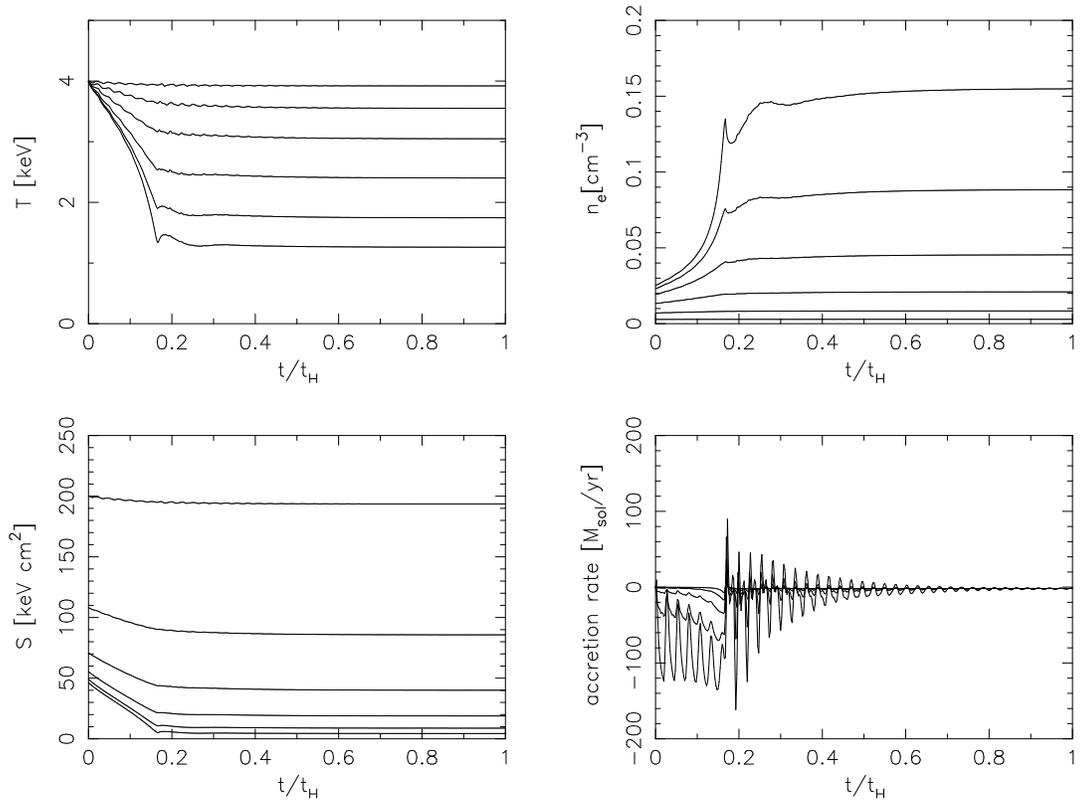}
\caption{The dependence of temperature (upper left panel), electron number density (upper
right), entropy (lower left) and accretion rate as a function of time for
different distances from the cluster center. The final accretion rate is
$\sim 1.76 M_{\odot}$ yr$^{-1}$. See text for details.}
\end{center}
\end{figure*}

\end{document}